\documentclass[twocolumn,trackchanges]{aastex631}

\usepackage{amsmath}
\usepackage{graphicx}
\usepackage{amsfonts}
\usepackage{natbib}
\usepackage{color}
\usepackage{epstopdf}
\usepackage{graphicx}
\usepackage{multirow}
\usepackage{longtable}
\usepackage{rotating}
\usepackage{xcolor}

\shorttitle{GeV $\gamma$-ray emission from  pulsar wind nebula G32.64+0.53}
\shortauthors{Xiao et al.}

\begin{document}

\title{Likely detection of GeV $\gamma$-ray emission from  pulsar wind nebula G32.64+0.53 with Fermi-LAT}

\author{Yifan Xiao}
\affiliation{Department of Astronomy, School of Physics and Astronomy, Key Laboratory of Astroparticle Physics of Yunnan Province, Yunnan University, Kunming 650091, People's Republic of China; fangjun@ynu.edu.cn}

\author{Keyao Wu}
\affiliation{Department of Astronomy, School of Physics and Astronomy, Key Laboratory of Astroparticle Physics of Yunnan Province, Yunnan University, Kunming 650091, People's Republic of China; fangjun@ynu.edu.cn}

\author[0000-0001-8043-0745]{Jun Fang}
\affiliation{Department of Astronomy, School of Physics and Astronomy, Key Laboratory of Astroparticle Physics of Yunnan Province, Yunnan University, Kunming 650091, People's Republic of China; fangjun@ynu.edu.cn}

\begin{abstract}

In this study, we report the likely GeV $\gamma$-ray emissions originating from the pulsar PSR J1849-0001's pulsar wind nebula (PWN) G32.64+0.53. Our analysis covers approximately 14.7 years of data from the Fermi Large Area Telescope (Fermi-LAT) Pass 8. The position of the source and its spectrum matches those in X-ray and TeV energy bands, so we propose that the GeV $\gamma$-ray source is indicative of PWN G32.64+0.53. We interpret the broadband spectral energy distribution (SED) using a time-dependent one-zone model, which assumes that the multi-band non-thermal emission of the target source can be generated by synchrotron radiation and inverse Compton scattering (ICS) of the electrons/positrons. Our findings demonstrate that the model substantially elucidates the observed SED. These results lend support to the hypothesis that the $\gamma$-ray source originates from the PWN G32.64+0.53 powered by PSR J1849-0001. Furthermore, the $\gamma$-rays in TeV bands are likely generated by electrons/positrons within the nebula through Inverse Compton Scattering.

\end{abstract}


\section{Introduction} \label{sec:intro}

Pulsars steadily dissipate their rotational energy through relativistic winds and transform the energy into magnetized ultrarelativistic particles, such as relativistic electrons and positrons. These particles formed PWN around pulsars, seen across the electromagnetic spectrum in synchrotron and inverse Compton emission \citep{PWN}. Pulsar wind particles are accelerated at the termination shock to hundreds of TeV or even PeV \citep{2023HEAD...2020107W}. Integrating theoretical models with data obtained from multi-wavelength observations allows for a comprehensive exploration of the primary contributors to Galactic cosmic rays (CRs). As an increasing number of observational facilities come into play, many PWNe are being identified. However, the research of GeV $\gamma$-ray PWN counterpart is still lacking, which makes many researchers search for GeV radiation of PWNe based on Fermi-LAT \citep{xiang,Straal,gong,Wu,wang}.

During a survey of the Sagittarius arm tangent region of the Galaxy, the INTEGTRAL observatory detected a source as IGR J18490-0000 \citep{2004first}. The source is centered on the position of PSR J1849-0001 and composed of a point-like source surrounded by an extended nebula, which suggests the presence of a pulsar wind nebula powered by PSR J1849-0001\citep{2008first}. PSR J1849-0001 is spinning down rapidly with period $P = 35~\mathrm{ms}$, yielding a spin-down luminosity $\dot{E} = 9.8 \times 10^{36}~\mathrm{erg\,s^{-1}}$, characteristic age $\tau_c \equiv P/2\dot{P} = 42.9~\mathrm{kyr}$, and surface dipole magnetic field strength $B_s = 7.5 \times 10^{11}~\mathrm{G}$\citep{2011pulsar}.

\begin{table*}
\centering
\label{tab:X-ray}
\caption{Photon index and flux measurements using different X-ray instruments.}
\begin{tabular}{lccr} 
\hline
\textbf{Instrument}  & \textbf{Photon Index} & \textbf{Flux ($\mathrm{erg \, cm}^{-2}\,\mathrm{s}^{-1}$)}& \textbf{Citation} \\ \hline
XMM-Newton  & $2.10 \pm 0.30$ & $(9.0 \pm 2.0) \times 10^{-13}$ & \citet{2011pulsar} \\
Chandra  & $1.18 \pm 0.05$ & $(7.1 \pm 0.5) \times 10^{-13}$ & \citet{2015kuiper} \\ 
XMM-Newton & $1.75 \pm 0.05$ & $(6.23 \pm 0.42) \times 10^{-13}$ & \citet{2015kuiper} \\ 
NuSTAR + Chandra & $2.25 \pm 0.29$ & $(13.7 \pm 1.5) \times 10^{-13}$ & \citet{2024ApJ...960...78K} \\ 
\hline
\end{tabular}
\end{table*}
The energy spectrum of PSR J1849-0001 and its PWN 32.64+0.5 have been studied in detail in both the X-ray and TeV energy bands. In X-ray band, the observational results of the PWN spectrum are as Table.~\ref{tab:X-ray}. The difference in photon index may be due to the different extraction regions and
methods used\citep{2015kuiper}. In the TeV band, HESS J1849-000 was associated with PSR J1849-0001. The emission of HESS J1849-000 spans from 320 GeV to 35 TeV and has a power-law spectrum with photon index of 2.39±0.08\citep{2018A&A...612A...1H}. The Tibet AS$\gamma$ array also detected the region of HESS J1849-000 in the energy range from 40 TeV to 320 TeV, described with a simple power-law function of $\mathrm{d}N/\mathrm{d}E = (2.86 \pm 1.44) \times 10^{-16} (E/40~\mathrm{TeV})^{-2.24 \pm 0.41}~\mathrm{TeV^{-1} cm^{-2} s^{-1}}$ \citep{2023ApJ...954..200A}. LHAASO \citep{2023arXiv230517030C} detected and calculated the TS value of sources above 100 TeV (denoted by ${TS}_{100}$). When ${TS}_{100} > 20$, the source was claimed as an Ultra High Energy (UHE) source. The UHE source 1LHAASO J1848-0001u was associated with IGR J18490-0000 and HESS J1849-000.

In GeV energy band, $\gamma$-ray radiation corresponding to the PWN is still lacking. With 3.6 yr of observations, the Fermi-LAT collaboration did not find any significant gamma radiation originating from region HESS J1849-000, which may because the HESS J1849-000 is located in regions of strong diffuse emission where the sensitivity of LAT is lower to any potential emission \citep{2013quiet}. Therefore, the lack of $\gamma$-ray-loud PWN cannot be proven given the current statistics and systematics associated with this analysis \citep{2013quiet}.

Since the GeV $\gamma$-ray emission of PSR J1849-0001 has not been analyzed in detail, we have carried out the study of the region near the source PSR J1849-0001 with 14.7 years of Fermi-LAT Pass 8 data here. The paper is organized as follows. In Section \ref{sec:data} we describe the data and data reduction. In Section \ref{sec:dataresult} we present the results of the data analysis. In Section \ref{sec:model} we analysis the broadband SED with one-zone leptonic model. In Section \ref{sec:summary} we discuss the results, and we conclude with a summary.

\section{Data Reduction} \label{sec:data}

The Fermi-LAT \citep{2009fermi}, a pair-conversion telescope, effectively detects $\gamma$-ray photons spanning from 20 MeV to well beyond 1 TeV. Operational since August 2008, this telescope conducts comprehensive all-sky surveys every 3 hours. The newest catalog of Fermi-LAT is the 4FGL-DR4 catalog, and the time for it was taken during the period 4 August 2008 (15:43 UTC) to 2 August 2022 (21:53 UTC) covering 14 years\citep{DR4}. In this catalog, the PSR J1849-0001 was not mentioned or detected. The last research for it by Fermi Collaboration only covered 45 months of data collected from 2008 August 4 to 2012 April 18 (Fermi mission elapsed time 239557440–356439741 s) for regions around the pulsar\citep{2013quiet}, which was much shorter than the running time of the Fermi-LAT.

In this study, in order to obtain more convincing data, we selected the Fermi-LAT Pass 8 Front+Back events, recorded from 2008 August 4 to 2023 April 9 (Fermi mission elapsed time 239557417s - 700012805s). The region of interest (ROI) is defined as a squared area of 20° × 20°, with its center positioned at the coordinates of PSR J1849-0001 (R.A.= 282.258°, Decl.=-0.022°). The energy spectrum considered spans from 10 to 500 GeV. We adopt the event class P8R3 SOURCE (“evclass = 128”) and event type FRONT + BACK (“evtype = 3”) as recommended by the LAT team. To ensure data quality, we filter the dataset based on the criteria (DATA\_QUAL $>$ 0) \&\& (LAT\_CONFIG == 1). A maximum zenith angle of 90° is set to minimize interference from the Earth Limb, and the instrumental response function used is “P8R3\_SOURCE\_V3”. For data processing, we employ a binning strategy of 0.1° per pixel for SED and 0.01° per pixel for TS map, with 8 bins per decade in energy. To model both Galactic emission and isotropic component backgrounds, we used the \texttt{gll\_iem\_v07.fits} template and the \texttt{iso\_P8R3\_V3\_v1.txt} template. To build our source model, we used the 4FGL-DR4 catalog, version \texttt{gll\_psc\_v32.fit}. During the fitting procedure with the binned analysis method, we set free
the spectral parameters and the normalizations of sources within
distance 3° to ROI center, as well as the normalizations of the two diffuse backgrounds.

The analysis is performed using fermitools v.2.2.0, with fermipy v1.1.6\citep{fermipy}.
\begin{figure*}
\centering

\begin{minipage}{0.45\textwidth}
\centering
\includegraphics[width=\linewidth]{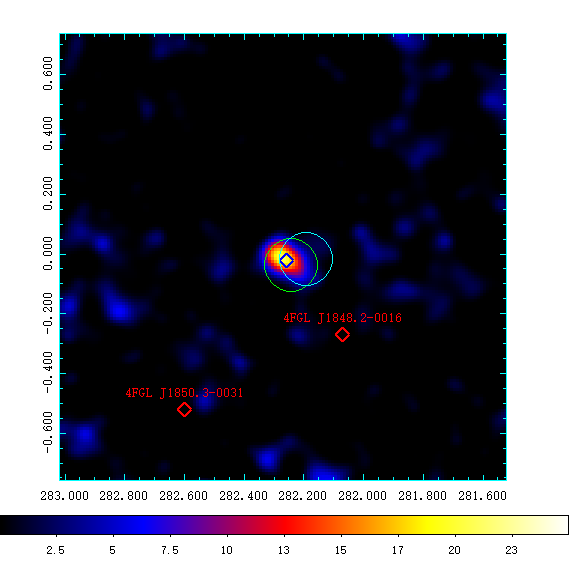} 
\end{minipage}
\begin{minipage}{0.45\textwidth}
\centering
\includegraphics[width=\linewidth]{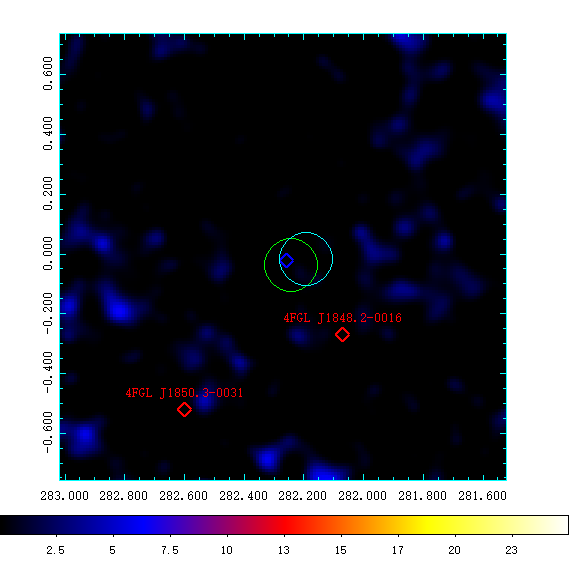}
\end{minipage}

\caption{Fermi-LAT TS map centered of SrcX region in the 10–500 GeV range. Left panel: TS map including all sources from 4FGL catalog. Right panel: TS map after subtracting all sources containing SrcX. The blue diamond represent the position of PSR J1849-0001. The red diamond represent the position of 4FGL-DR4 sources. The positions of 1LHAASO J1848-0001u and HESS J1849-000 are marked with cyan and green circles.}

\label{fig:TS}  
\end{figure*}
\begin{table*}
\centering
\caption{Spatial Distribution Analysis for SrcX with Two Types of Spatial Models in 10–500 GeV}
\label{tab:ts}
\begin{tabular}{c c c c c c} 
\hline
Spatial Template & Radius($\sigma$) & Photon Index & Photon Flux & TS Value & Degrees of Freedom\\
& degree &  & $\mathrm{10}^{-11}$ ph $\mathrm{cm}^{-2}$ $\mathrm{s}^{-1}$ &   &  \\
\hline
Point source & … & 2.35$\pm$0.39 & 4.35$\pm$1.42& 19.48 & 4 \\
uniform disk & 0.03& 2.35$\pm$0.38 & 4.52$\pm$1.46 & 18.42 & 5 \\
2D Gaussian & 0.03& 2.34$\pm$0.37 & 4.54$\pm$1.45 & 18.43 & 5 \\
\hline
\end{tabular}
\end{table*}
\section{results} \label{sec:dataresult}

\subsection{Spatial Analysis}\label{subsec:pointsource}
To investigate whether there is a detection of PSR J1849-0001 and PWN G32.64+0.53, we firstly consider about the nearby 4FGL sources. However, in region around the pulsar, the nearest 4FGL-DR4 source is 4FGL J1848.2-0016, described as a point source with a Log-Parabola spectrum. We do not consider this source to be the counterpart, as it is at 0.31° away from the pulsar and not covered by the detection of H.E.S.S. or LHAASO.

To achieve the best-fit model, we compute a test statistic (TS) map. The TS is formed as : TS = 2(ln$\mathcal{L}_1$ - ln $\mathcal{L}_0$), where $\mathcal{L}_1$ is the maximum likelihood for a model including a new source and $\mathcal{L}_0$ is the maximum likelihood for a model without the source. As shown in Fig.~\ref{fig:TS}, we have established a 1.5 ° × 1.5 ° TS map and find significant $\gamma$-ray radiation with TS = 19.48 from the direction of PSR J1849-0001. After subtracting the point source as well, there is no other significant residual $\gamma$-ray radiation from this location. The best-fit position (R.A.= 282.275°,Decl.=-0.017°) was calculated by running the $\texttt{find\_source}$ of fermipy. The new $\gamma$-ray source was marked as SrcX.

To check whether SrcX is an extended source, we tested the source for extension using the extension templates RadialDisk and RadialGaussian. The fitting results are shown in Table.~\ref{tab:ts}. Both these extensions are not more significantly than the point source, so we conclude that there is no evidence for source extension. We used the point-source template to do all subsequent analyses.

\subsection{Variability Analysis}\label{subsec:lightcurve}
\begin{figure*}
\centering
\includegraphics[width=0.8\textwidth]{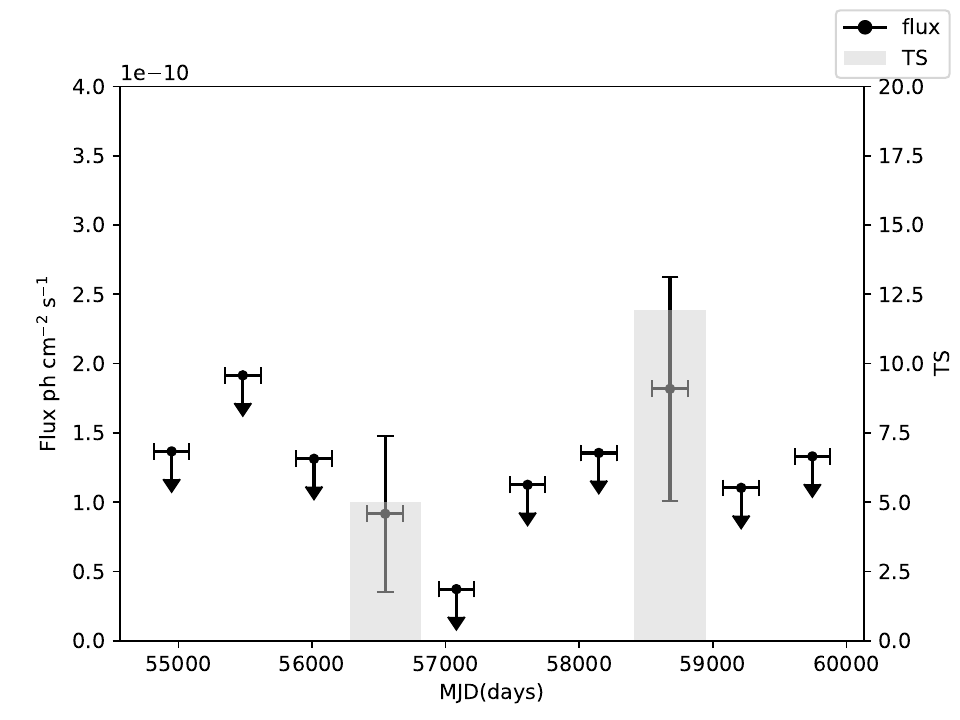}
    \caption{Light curve of SrcX of 10 to 500 GeV for 10 time bins. The time bin is set to the upper limits with the 95\% confidence level when TS value $<$ 4. The gray histogram is used to mark the TS value for each time bin with a TS value $>$ 4.}
    \label{fig:lc}
\end{figure*}

We tested the possible variability of the SrcX by constructing a light curve.
We divided the 14.7 year data (MJD 54682 to 60012) into 10 time bins in the 10-500 GeV band as Fig.~\ref{fig:lc}, for the light curve, ${TS}_{var}$ $>$ 23.1 is used to identify variable sources at a 99\%
confidence level\citep{TSvar}. We find a ${TS}_{var}$ = 7.34 for SrcX, which indicates no significant variability for the $\gamma$-ray emission from SrcX.\\
\subsection{Spectral Analysis\label{subsec:GeVsed}}
The $\gamma$-ray emission signal we detected is only at $\sim$4$\sigma$ confidence level, which is lower than most of the confirmed $\gamma$-ray pulsars listed in 4FGL. According to the SED of $\gamma$-ray pulsar provided by Fermi-LAT\citep{PSR3}, the radiation of pulsars above 10GeV is usually not visible, so we assume that all observed radiation comes from PWN G32.64+0.53.

We selected the energy band of 10–500 GeV to generate its spectral energy distribution and divided the energy range into 6 bins. If the TS value of energy bin is less than 4.0, then an upper limit with 95\% confidence level needs to be calculated. The spectrum is best described by a power law model formed as follows:
\begin{equation}
    \frac{dN}{dE} = N_0 \left(\frac{E}{E_0}\right)^{-\Gamma},
        \label{eq:powerlaw}
\end{equation}
In which, the photon index $\Gamma = 2.35\pm0.39$, the prefactor $N_0 = (1.158 \pm 0.379) \times 10^{-15} \, \text{cm}^{-2}\text{s}^{-1}\text{MeV}^{-1}$, and the energy scale $E_0$ is fixed to 20000 MeV.

\begin{figure}
\includegraphics[width=\columnwidth]{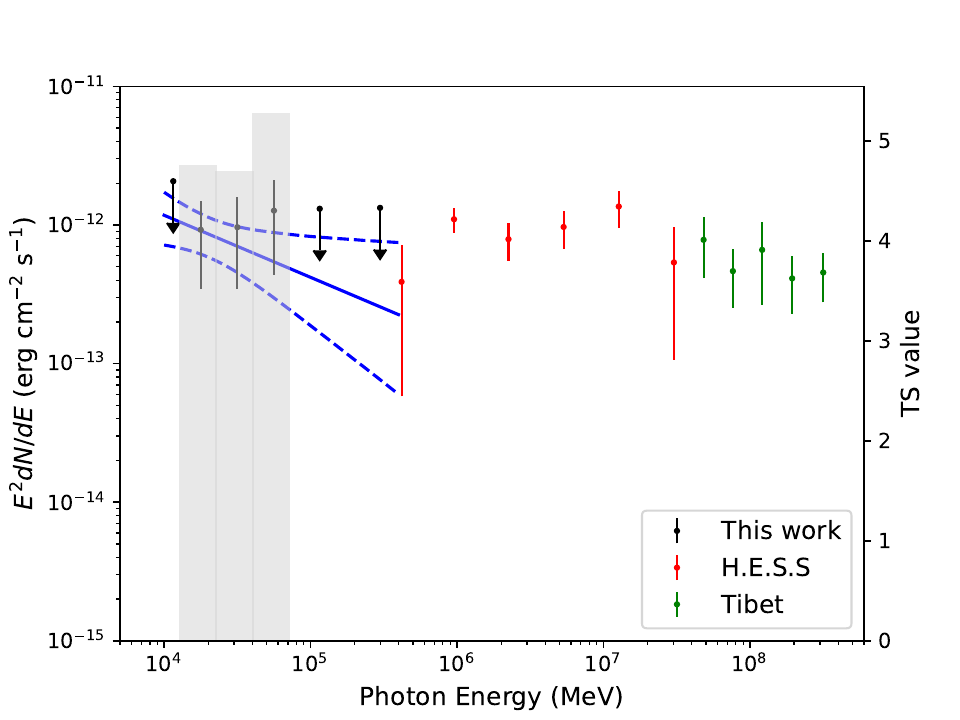}
    \caption{The Fermi-LAT spectral data analysis results (marked as black
dots) of SrcX in the 10–500 GeV band. The gray histogram represents the
TS value for each energy bin. A power-law spectrum with an index of 1.94
$\pm$ 0.28 for the GeV data is plotted by the blue solid line, and its 1$\sigma$ statistical error is represented by the blue dashed lines. The red and green dots display the TeV $\gamma$-ray data observed by HESS\citep{2018A&A...612A...1H} and AS$\gamma$\citep{2023ApJ...954..200A}. }
    \label{fig:fermised}
\end{figure}

To investigate the systematic errors in the local $\gamma$-ray Galactic background and the effective area, we altered the normalization of the diffuse background by ±5\% and use the isotropic diffuse background version P8R3\_SOURCE\_V2 instead of P8R3\_SOURCE\_V3\citep{2013ApJ...773...77A}.
The spectral points we obtained are shown in Fig.~\ref{fig:fermised}, which is well connected to the TeV spectrum of HESS J1849-000 and AS$\gamma$. The best-fit data from the 6 energy bins, accounting for the systematic uncertainties, are presented in Table.~\ref{tab:sed}.
\begin{table}
\centering
\caption{Observed Energy Flux of SrcX with Fermi-LAT}
\label{tab:sed}
\begin{tabular}{lccr} 
\hline
Energy & Band & $\mathrm{E}^{2}$dN(E)/dE & TS value\\
GeV & GeV & $\mathrm{10}^{-13}$ ergs $\mathrm{cm}^{-2}$ $\mathrm{s}^{-1}$ & \\
\hline
11.5 & 10-13.3 & $<$20.6 & 2.46\\
17.8 & 13.3-23.7 & $(4.05-13.9)_{-0.60}^{+1.12}$ & 4.76\\
31.6 & 23.7-42.2 & $(3.89-15.1)_{-0.43}^{+0.74}$ & 4.69\\
56.2 & 42.2-75.0 & $(4.67-20.4)_{-0.33}^{+0.63}$ & 5.28\\
115.5 & 75.0-177.8 & $<$13.1 & 0.01\\
298.1 & 177.8-500.0 & $<$13.3 & 0\\
\hline
\end{tabular}
\end{table}

\section{Modeling of Pulsar Wind Nebula} \label{sec:model}
\begin{figure}
\includegraphics[width=\columnwidth]{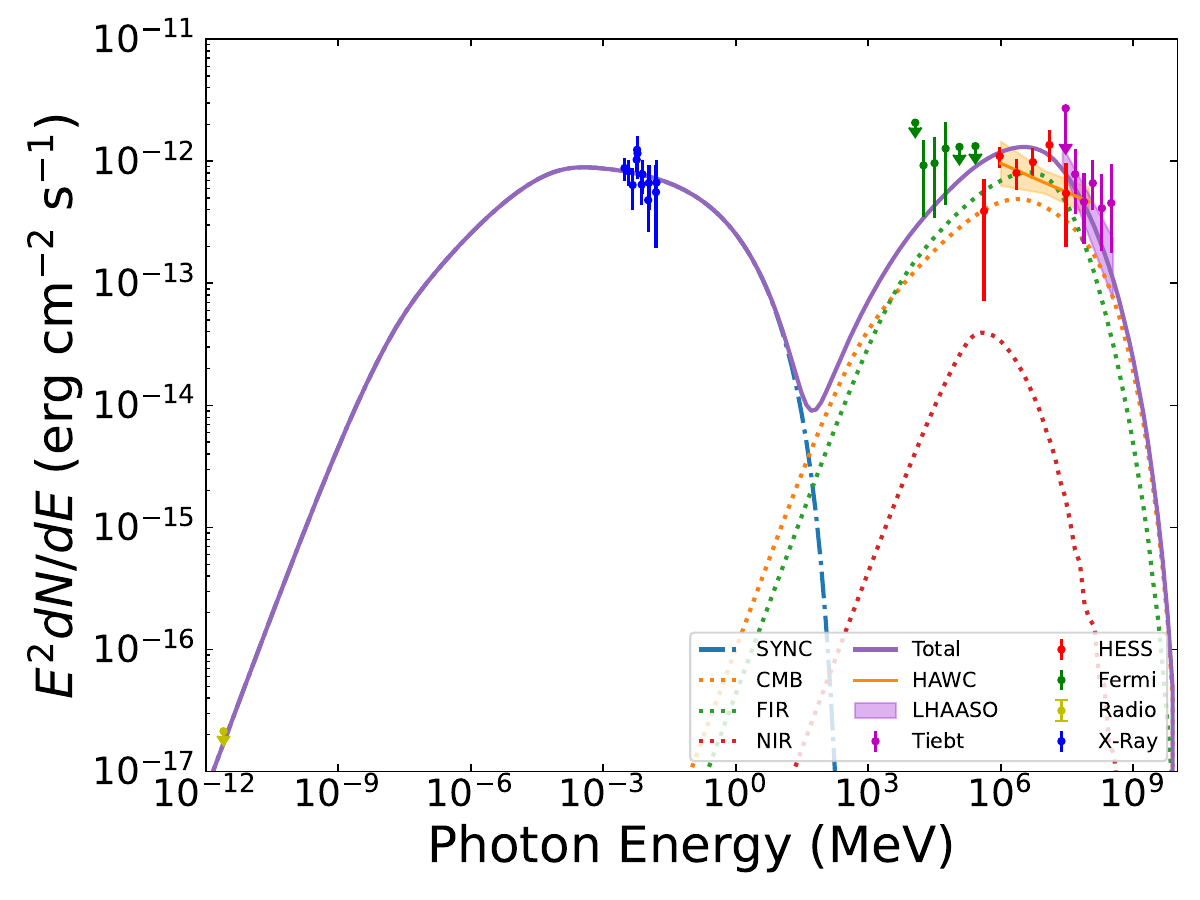}
    \caption{Comparing the SED derived from the time-dependent one-zone model with the detected fluxes associated with LHAASO J1848-0001u, the total emission (solid line) incorporates contributions from three considered photon fields for ICS, and includes the predicted synchrotron radiation from a 3.69$\mu$G magnetic field. The green data points represent the results of this work, while other data points are sourced from HESS\citep{2018A&A...612A...1H}, HAWC\citep{2022icrc.confE.827H}, LHAASO\citep{2023arXiv230517030C}, Tibet AS$\gamma$\citep{2023ApJ...954..200A} and X-ray data obtained from Chandra and NuSTAR\citep{2024ApJ...960...78K}. The observation conducted with Giant Metrewave Radio Telescope at 610 MHz revealed a radio flux density upper limit of 3.5 mJy for LHAASO J1848-0001u\citep{2006A&A...453...83P}.}
    \label{fig:1849sed}
\end{figure}

In this part, the findings of the study shed light on the origin of $\gamma$-ray emission from LHAASO J1848-0001u and its association with PWN G32.64+0.53. Observed $\gamma$-rays from HESS J1849-000 using the Tibet AS$\gamma$\, with significances reaching 4.0$\sigma$ at 25 TeV and 4.4$\sigma$ at energies exceeding 100 TeV. The study described the energy spectrum for 40 TeV $<$ E $<$ 320 TeV\citep{2023ApJ...954..200A}. Previous studies on LHAASO J1848-0001u demonstrated discrepancies in spectral fitting, some employing leptonic models, while others used hadronic models. Additionally, some proposed that it might be a PWN-SNR composite system. However, this study supplemented Fermi measurement results with previously published multi-wavelength data and investigated the radiation characteristics of the source based on the one-zone leptonic model. It was found that LHAASO J1848-0001u is associated with a PWN powered by the pulsar PSR J1849-0001. Consequently, there is a preference for the origin of $\gamma$- ray emission from this source being linked to leptons, with multi-band non-thermal emission being generated through synchrotron radiation and ICS. We obtained a broadband SED as shown in Fig.~\ref{fig:1849sed}.\\
In this study, we employ a time-dependent one-zone model to simulate the radiative evolution of a PWN. We assume that the emission region of the $\gamma$-rays source is a sphere containing high-energy electrons/positrons accelerated by the termination shock of a PWN.

The energy distribution of particles at time $t$ can be derived from the equation\citep{2012MNRAS.427..415M,2010A&A...515A..20F,2014JHEAp...1...31T}.
\begin{equation}
   \frac{\partial N(\gamma,t)}{\partial t}=-\frac{\partial}{\partial \gamma}[\Dot{\gamma}(\gamma,t)N(\gamma,t)]-\frac{N(\gamma,t)}{\tau(\gamma,t)}+Q(\gamma,t) ,
\end{equation} 

The left side of the equation represents how the particle distribution changes over time. The first term on the right-hand side accounts for the continuous energy change in particles due to energy losses. The function $\Dot{\gamma}(\gamma,t)$ is the sum of energy loss rate due to synchrotron radiation, ICS, and adiabatic expansion. $Q(\gamma,t)$ represents the injection of particles per unit energy per unit time and $\tau(\gamma,t)$ represents the escape time. High-energy leptons are injected into the PWN following a broken power-law distribution, \citep{2004A&A...422..609B,2010A&A...515A..20F,2012MNRAS.427..415M,2014JHEAp...1...31T}
\begin{equation}
   Q(\gamma,t)= Q_0(t)
   \begin{cases}
   (\frac{\gamma}{\gamma_\mathrm{b}})^{\mathrm{-\alpha_1}} \quad if \gamma \le \gamma_\mathrm{b}\\
   (\frac{\gamma}{\gamma_\mathrm{b}})^{\mathrm{-\alpha_\mathrm{2}}} \quad if \gamma_\mathrm{b} < \gamma \le \gamma_\mathrm{max}
   \end{cases},
\end{equation}
where $\gamma_\mathrm{b}$ represents the break energy, while $\alpha_1$ and $\alpha_2$ denote the spectral indices. In our analysis, we assume that particles are injected at all Lorentz factors greater than 1. The normalization constant $Q_0(t)$ for the injection function can be determined as follows:
\begin{equation}
   (1-\eta) L(t)=\int\gamma m_e c^2 Q(\gamma,t) d\gamma ,
\end{equation}
Additionally, $\gamma_\mathrm{max}$ is the super-exponential cutoff energy, reflecting the truncation characteristic of the particle energy spectrum. To ensure that particles are confined within the termination shock of PWN, a specific condition must be satisfied by\citep{2009ASSL..357..451D}:
\begin{equation}
   \gamma_{\mathrm{max}}=\frac{\epsilon e \kappa}{m_{\mathrm{e }}c^2}\sqrt{\eta\frac{L(t)}{c}} ,
\end{equation}
where $\epsilon$ represents the fractional size of the shock radius, which must be less than 1, while $\eta$ denotes the particle energy fraction, and $\kappa$ represents the magnetic compression ratio. Additionally, $L(t)$ represents the spin-down luminosity of a pulsar, while $L_0$ denotes the initial luminosity. This can be expressed as follows\citep{1969ApJ...157..869G,2006ARA&A..44...17G}:
\begin{equation}
   L(t)=4\pi^2 I\frac{\dot{P}}{P^3}=L_0(1+\frac{t}{\tau_0})^{-\frac{n+1}{n-1}} ,
\end{equation}

\begin{table*}
    \centering
    \caption{Fitting parameters of the PWN model.}
    \label{tab:table4}
    \setlength{\tabcolsep}{31pt} 
    \begin{tabular}{llll}     
        \hline
        Magnitude & Symbol & Value & Comment \\
        \hline
        CMB temperature (K) & \( T_{\mathrm{CMB}} \) & 2.73 & Fixed\\
        CMB energy density (eV/cm\(^3\)) & \( U_{\mathrm{CMB}} \) & 0.25 & Fixed\\
        FIR temperature (K) & \( T_{\mathrm{FIR}} \) & 20 & Fixed\\
        FIR energy density (eV/cm\(^3\)) & \( U_{\mathrm{FIR}} \) & 0.75 & Fixed \\
        NIR temperature (K) & \( T_{\mathrm{NIR}} \) & 3000 & Fixed\\
        NIR energy density (eV/cm\(^3\)) & \( U_{\mathrm{NIR}} \) & 1.26 & Fixed\\
        Ejected mass (\( M_{\odot} \)) & \( M_{\mathrm{ej}} \) & 9.5 &  Fixed\\
        SN explosion energy (erg) & \( E_0 \) & \( 10^{51} \) &Fixed \\
        Age (yr) & \( t_{\mathrm{age}} \) & 21000 & Result of the fit\\
        Minimum energy at injection & \( \gamma_{min} \) & 1 & Assumed \\
        Low energy index & \( \alpha_1 \) & 1.0 & Result of the fit \\
        High energy index & \( \alpha_2 \) & 2.0 & Result of the fit\\
        Initial spin-down time-scale (yr) & \( \tau_0 \) & 5000 & Result of the fit\\
        Magnetic field strength (\( \mu \mathrm{G} \)) & \( B \) & 3.69 & Result of the fit\\
        Magnetic fraction & \( \eta \) & 0.3 & Result of the fit\\
        Break energy & \( \gamma_b \) & \( 6 \times 10^7 \) & Result of the fit\\
        Shock radius fraction& \( \epsilon \) &  0.4 & Result of the fit\\
        \hline
    \end{tabular}
\end{table*}

where $P$ and $\dot{P}$ represent the period and its first derivative, respectively, and $n$ represents the braking index of the pulsar. The moment of inertia of a pulsar, defined as $I$, is assumed to be approximately $10^{45} \mathrm{g} \,\mathrm{cm}^2$. The expression for the initial spin timescale $\tau_0$ of a pulsar can be written as follows\citep{2006ARA&A..44...17G}:
\begin{equation}
   \tau_0=\frac{P_0}{(n-1)\dot{P_0}}=\frac{2\tau_c}{n-1}-t_{\mathrm{age}} .
\end{equation}
where $P_0$ represents the initial period, $\dot{P_{\mathrm{0}}}$ represents its initial first derivative, and $\tau_{\mathrm{c}} = P/2\dot{{P}}$ represents the characteristic age of the pulsar.

The expansion dynamics of PWN within SNR are categorized into three distinct phases, determined by the system age t, the initial spin-down timescale $\tau_0$, and the reverse-shock interaction time $t_{\text{rs}}$. This relationship can be expressed as follows\citep{2012arXiv1202.1455M,2018A&A...612A...2H}:

\begin{equation}
     R(t) \propto \begin{cases} 
     t^{6/5} & \mathrm{for} \, t \leq \tau_0, \\
     t & \mathrm{for} \, \tau_0 < t \leq t_{\mathrm{rs}}, \\
     t^{3/10} & \mathrm{for} \, t > t_{\mathrm{rs}}.
     \end{cases},
\end{equation}
To simplify the model, we did not self-consistently simulate the compression and re-expansion by the reverse shock (RS) as done in other works\citep{2022ApJ...930..148B,2022ApJ...940..143E,2023ApJ...945....4E,2023ApJ...946...40A,2023HEAD...2020107W,2024ApJ...960...75P}, but instead chose not to model the crushing of the PWN by the SNR reverse shock. The magnetic field strength in the PWN is\citep{2012MNRAS.427..415M}
\begin{equation}
    B(t) = \sqrt{\frac{3(n-1) \eta L_{\mathbf{0}} \tau_{\mathrm{0}}} {R_{\mathrm{PWN}}^{3}(t)} \left[ 1-\left(1+\frac{t}{\tau_{\mathrm{0}}}\right)^{-\frac{2}{n-1}}\right] } ,
\end{equation}

In our model, we have considered three crucial interstellar photon fields and their respective energy densities. Firstly, the cosmic microwave background (CMB) possesses a temperature of $T_{\mathrm{CMB}} = 2.73$ K and an energy density of $U_{\mathrm{CMB}} = 0.25 \,\mathrm{eV} \,\mathrm{cm}^{-3}$. Secondly, the infrared (IR) radiation is separated into two parts,  far-infrared(FIR) component exhibits a temperature of $T_{\mathrm{FIR}} = 20.0$ K with an energy density of $U_{\mathrm{FIR}} = 0.75 \,\mathrm{eV} \,\mathrm{cm}^{-3}$, while the near-infrared component has a temperature of $T_{\mathrm{NIR}} = 3000.0$ K with an energy density of $U_{\mathrm{NIR}} = 1.26 \,\mathrm{eV} \,\mathrm{cm}^{-3}$\citep{2017ApJ...846...67P}. In addition, we also consider the kinetic energy of the supernova ejecta is $E_0 = 10^{51} \,\mathrm{erg}$ with a ejected mass of $M_{\mathrm{ej}} = 9.5 M_{\odot}$.

In this work we take energy index$\alpha_1$ and $\alpha_2$, the break energy  $\gamma_b$, the shock radius fraction $\epsilon$ , magnetic fraction $\eta$ and initial spin-down time-scale $\tau_0$ as the free parameters to reproduce SEDs consistent with multi-wavelength data. We assumed that the nebula has an age of 21 kyr in order to make the synchrotron spectra coincide with the observations of x-ray data.
And we found that the detected multi-wavelength data can be well reproduced by combining the parameters $\alpha_1 = 1.0$, $\alpha_2 = 2.0$, $\gamma_b=6.0 \times 10^7$, $\eta=0.3$, $\epsilon=0.4$, and $\tau_0=5000$. The fitting parameters used for our model are summarized in Table.~\ref{tab:table4} and the resulting SED is shown in Fig.~\ref{fig:1849sed}. Furthermore, in comparison with other research, we found that \citep{2024ApJ...960...78K} used a multi-zone emission model to determine a lower magnetic field strength within the PWN, approximately $7 \,\mu$G. However, based on our research model and constrained by the X-ray flux, the calculated magnetic field is even lower, estimated at around $3.69 \,\mu$G. This is significantly lower than the magnetic field strength observed in the Crab Nebula, which is estimated to be about $100 \,\mu$G\citep{2012MNRAS.427..415M,2014JHEAp...1...31T,2021RAA....21..286W,2022JHEAp..36..128M}. As postulated by the model advanced by \citep{2009ApJ...703.2051G}, the magnetic field intensity within PWNe is subject to a gradual diminution as they undergo continuous evolutionary processes within SNR. While subject to fluctuations, this attenuation trend culminates in a significant reduction of the magnetic field strength to several to 10 $\mu$G upon the PWN reaching an approximate age of 10 kyr.

\section{Conclusions} \label{sec:summary}
In this study, by analysis on $\sim$14.7 years of data from the Fermi Large Area Telescope (Fermi-LAT) Pass 8, we get the TS map, light curve and SED of PWN G32.64+0.53. The TS map shows a $\sim$4$\sigma$ significance $\gamma$-ray emission located on the position of G32.64+0.53. No significant variability in the light curve is found with ${TS}_{var}$ = 7.34 significance. The GeV SED matches those in X-ray and TeV energy bands, so we propose that the GeV $\gamma$-ray source is the GeV counterpart of PWN G32.64+0.53. Combining our GeV $\gamma$-ray data, we investigate the multi-band non-thermal radiation properties of the PWN G32.64+0.53 with the time-dependent one-zone model. The fitting result of the model is consistent with the multi-band fluxes with appropriate parameters. This indicates that high-energy $\gamma$-ray of PWN G32.64+0.53 may originate from the leptonic process of electrons/positrons via ICS. LHAASO J1848-0001u is also associated with the PWN powered by PSR J1849-0001.

\begin{acknowledgments}
This research is supported by NSFC through Grants 12063004 and 12393852, as well as grants from the Yunnan Provincial Government (YNWR-QNBJ-2018-049) and Yunnan Fundamental Research Projects (grant No. 202201BF070001-020). Y.F.X and K.Y.W are supported by Scientific Research and Innovation Project of Postgraduate Students in the Academic Degree of Yunnan University under grant KC-23233697 and KC-23233957.
\end{acknowledgments}

\bibliography{sample631}{}
\bibliographystyle{aasjournal}

\end{document}